\documentclass[final]{svjour3}
\usepackage{graphicx}
\usepackage{rotating}
\usepackage{amssymb}
\usepackage{mathptmx}
\usepackage[numbers]{natbib}
\usepackage{verbatim}
\makeatletter
\journalname{Journal of Low Temperature Physics}
\bibpunct{}{}{,}{s}{}{,}

\begin{document}

\newcommand{\hdblarrow}{H\makebox[0.9ex][l]{$\downdownarrows$}-}
\title{Plastic Laminate Antireflective Coatings for Millimeter-wave Optics in BICEP Array}

\author{M.~Dierickx$^1$ \and P.~A.~R.~Ade$^2$ \and Z.~Ahmed$^{3,4}$ \and M.~Amiri$^5$ \and D.~Barkats$^1$ \and R.~Basu Thakur$^6$ \and C.~A.~Bischoff$^7$ \and D. Beck$^{3,4}$ \and J.~J.~Bock$^{6,8}$ \and V.~Buza$^1$ \and J.~Cheshire$^9$ \and J.~Connors$^1$ \and J.~Cornelison$^1$ \and M.~Crumrine$^9$ \and A.~Cukierman$^{3,4}$ \and E. Denison$^{13}$ \and L.~Duband$^{10}$ \and M.~Eiben$^1$ \and S.~Fatigoni$^5$ \and J.~P.~Filippini$^{11,12}$ \and N.~Goeckner-Wald$^4$ \and D.~C.~Goldfinger$^1$ \and J.~A.~Grayson$^4$ \and P.~Grimes$^1$ \and G.~Hall$^{1,9}$ \and G. Halal$^7$ \and M.~Halpern$^5$ \and E.~Hand$^5$ \and S.~Harrison$^1$ \and S.~Henderson$^3$ \and S.~R.~Hildebrandt$^8$ \and G.~C.~Hilton$^{13}$ \and J. Hubmayr $^13$ \and H.~Hui$^6$ \and K.~D.~Irwin$^{3,4}$ \and J.~Kang$^4$ \and K.~S.~Karkare$^{1,14}$ \and S.~Kefeli$^6$ \and J.~M.~Kovac$^1$ \and C.~L.~Kuo$^4$ \and K.~Lau$^9$ \and E.~M.~Leitch$^{14}$ \and A. Lennox$^11$ \and K.~G.~Megerian$^8$ \and L.~Minutolo$^6$ \and L.~Moncelsi$^6$ \and Y.~Nakato$^4$ \and T.~Namikawa$^{15}$ \and H.~T.~Nguyen$^{6,8}$ \and R.~O'Brient$^{6,8}$ \and S.~Palladino$^7$ \and M.~Petroff$^1$ \and N.~Precup$^9$ \and T.~Prouve$^{10}$ \and C.~Pryke$^9$ \and B.~Racine$^1$ \and C.~D.~Reintsema$^{13}$ \and D.~Santalucia$^1$ \and A.~Schillaci$^6$ \and B.~L.~Schmitt$^1$ \and B.~Singari$^9$ \and A.~Soliman$^6$ \and T.~St.~Germaine$^1$ \and B.~Steinbach$^6$ \and R.~V.~Sudiwala$^2$ \and K.~L.~Thompson$^4$ \and C.~Tucker$^2$ \and A.~D.~Turner$^8$ \and C.~Umilt\`{a}$^7$ \and C.~Verges$^1$ \and A.~G.~Vieregg$^{14}$ \and A.~Wandui$^6$ \and A.~C.~Weber$^8$ \and D.~V.~Wiebe$^5$ \and J.~Willmert$^9$ \and W.~L.~K.~Wu$^{14}$ \and E.~Yang$^{3,4}$ \and K.~W.~Yoon$^4$ \and E.~Young$^4$ \and C.~Yu$^4$ \and L.~Zeng$^1$ \and C.~Zhang$^6$ \and S.~Zhang$^6$}

\institute{\email{mdierickx@cfa.harvard.edu}
\\$^1$Center for Astrophysics, Harvard-Smithsonian, Cambridge, Massachusetts 02138, USA
\\$^2$School of Physics and Astronomy, Cardiff University, Cardiff, CF24 3AA, United Kingdom
\\$^3$Kavli Institute for Particle Astrophysics and Cosmology, SLAC National Accelerator Laboratory, 2575 Sand Hill Rd, Menlo Park, California 94025, USA
\\$^4$Department of Physics, Stanford University, Stanford, California 94305, USA
\\$^5$Department of Physics and Astronomy, University of British Columbia, Vancouver, British Columbia, V6T 1Z1, Canada
\\$^6$Department of Physics, California Institute of Technology, Pasadena, California 91125, USA
\\$^7$Department of Physics, University of Cincinnati, Cincinnati, Ohio 45221, USA
\\$^8$Jet Propulsion Laboratory, Pasadena, California 91109, USA
\\$^9$Minnesota Institute for Astrophysics, University of Minnesota, Minneapolis, 55455, USA
\\$^{10}$Service des Basses Temp\'{e}ratures, Commissariat \`{a} l'Energie Atomique, 38054 Grenoble, France
\\$^{11}$Department of Physics, University of Illinois at Urbana-Champaign, Urbana, Illinois 61801, USA
\\$^{12}$Department of Astronomy, University of Illinois at Urbana-Champaign, Urbana, Illinois 61801, USA
\\$^{13}$National Institute of Standards and Technology, Boulder, Colorado 80305, USA
\\$^{14}$Kavli Institute for Cosmological Physics, University of Chicago, Chicago, IL 60637, USA
\\$^{15}$Department of Applied Mathematics and Theoretical Physics, University of Cambridge, Wilberforce Road, Cambridge CB3 0WA, UK
\\$^{16}$Department of Physics, University of Toronto, Toronto, Ontario, M5S 1A7, Canada
\\$^{17}$Canadian Institute for Advanced Research, Toronto, Ontario, M5G 1Z8, Canada
\\$^{18}$Physics Department, Brookhaven National Laboratory, Upton, NY 11973}

\maketitle

\begin{abstract}

The BICEP/\textit{Keck} series of experiments target the Cosmic Microwave Background at degree-scale resolution from the South Pole. Over the next few years, the “Stage-3” BICEP Array (BA) telescope will improve the program's frequency coverage and sensitivity to primordial B-mode polarization by an order of magnitude. The first receiver in the array, BA1, began observing at 30/40 GHz in early 2020. The next two receivers, BA2 and BA3, are currently being assembled and will map the southern sky at frequencies ranging from 95~GHz to 150~GHz. Common to all BA receivers is a refractive, on-axis, cryogenic optical design that focuses microwave radiation onto a focal plane populated with antenna-coupled bolometers. High-performance antireflective coatings up to 760 mm in aperture are needed for each element in the optical chain, and must withstand repeated thermal cycles down to 4 K. Here we present the design and fabrication of the 30/40~GHz anti-reflection coatings for the recently deployed BA1 receiver, then discuss laboratory measurements of their reflectance. We review the lamination method for these single- and dual-layer plastic coatings with indices matched to various polyethylene, nylon and alumina optics. We also describe ongoing efforts to optimize coatings for the next BA cryostats, which may inform technological choices for future Small-Aperture Telescopes of the CMB “Stage 4” experiment.


\end{abstract}

\section{Introduction}

Since the discovery of the Cosmic Microwave Background (CMB) in 1965, precision observations have provided evidence in support of a Big Bang origin of the Universe. Small anisotropies in background temperature trace density (scalar) perturbations that subsequently led to the growth of large-scale structure. Since 2006 the BICEP/\textit{Keck} (BK) series of experiments has mapped CMB polarization from its observing site at the South Pole. The curl component of the polarization, also known as $B$-modes, may carry a faint signal caused by primordial gravitational waves. Such tensor perturbations naturally arise in theories of cosmological inflation describing an early phase of rapid expansion in the first $10^{-32}$~s after the Big Bang. Today the key to constraining inflationary models (generally parametrized by the the tensor-to-scalar ratio $r$) is disentangling the primordial $B$-mode signal from astrophysical foregrounds and gravitational lensing. Data taken by the BICEP2, \textit{Keck Array}, and BICEP3 CMB polarization experiments up to and including the 2018 observing season presently limit $r$ to $< 0.036$ at $95\%$ confidence [1].

\begin{figure}[htbp]
	\begin{center}
		\includegraphics[width=0.7\linewidth, keepaspectratio]{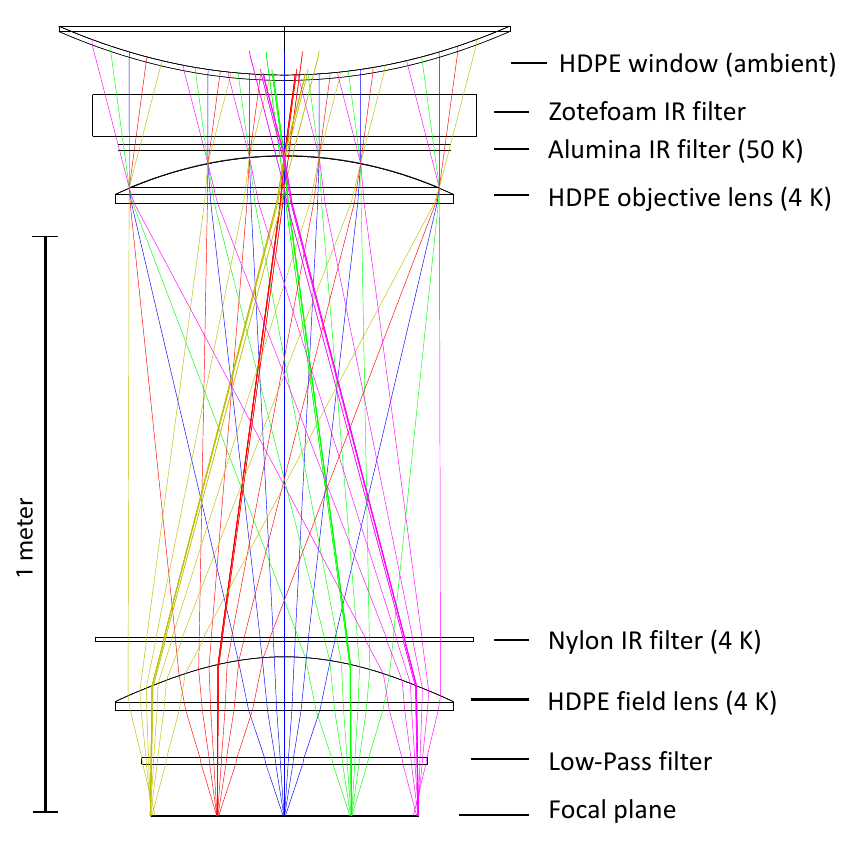}
		\caption{Schematic of optical elements in the BA1 receiver. The vacuum window and both lenses are made of HDPE. The design includes an alumina infrared filter held at 50~K and a nylon filter at 4~K. For reference, the lens and window diameters are 638~mm and 876~mm, respectively. See [3] for further details on the BA1 instrument. (Color figure online.)}
		\label{fig_BA_optics}
	\end{center}
\end{figure}

BICEP Array (BA) is a third-generation refractive telescope that consists of 4 receivers observing at microwave frequencies ranging from 30~GHz (deployed in 2019) to 270~GHz (in development) [2,3]. Most of the optical chain (presented in Fig.~\ref{fig_BA_optics}) is held at cryogenic temperatures in order to reduce thermal loading on its superconducting detectors. Materials are chosen for low loss at microwave frequencies combined with high absorption in the infrared: High-Density Polyethylene (HDPE) for the lenses (held at 4~K) and the ambient-temperature vacuum window; alumina and nylon for infrared filters at 50~K and 4~K, respectively. Optical elements are coated with antireflective (AR) layers in order to mimize stray reflections previously seen to cause systematic effects in the beam response. AR coatings must produce low reflected power over observational bandpasses, and withstand differential thermal contraction during cryogenic cycles.

In these proceedings we present the development of laminated antireflective coatings for single-band observations at  30/40~GHz, 95~GHz and 150~GHz. Section~\ref{sec_alumina} outlines the production and testing of custom epoxy AR layers for alumina substrates, in this case the 50~K alumina filter of the first BICEP Array receiver (known as BA1). Section~\ref{sec_plastics} presents the fabrication method and performance of AR coatings designed for HDPE and nylon optics at 30/40~GHz (deployed in BA1), 95~GHz and 150~GHz (in development for future CMB receivers).

\section{Epoxy AR-Coatings for Alumina}
\label{sec_alumina}

\begin{figure}[htbp]
\begin{center}
\includegraphics[width=0.7\linewidth, keepaspectratio]{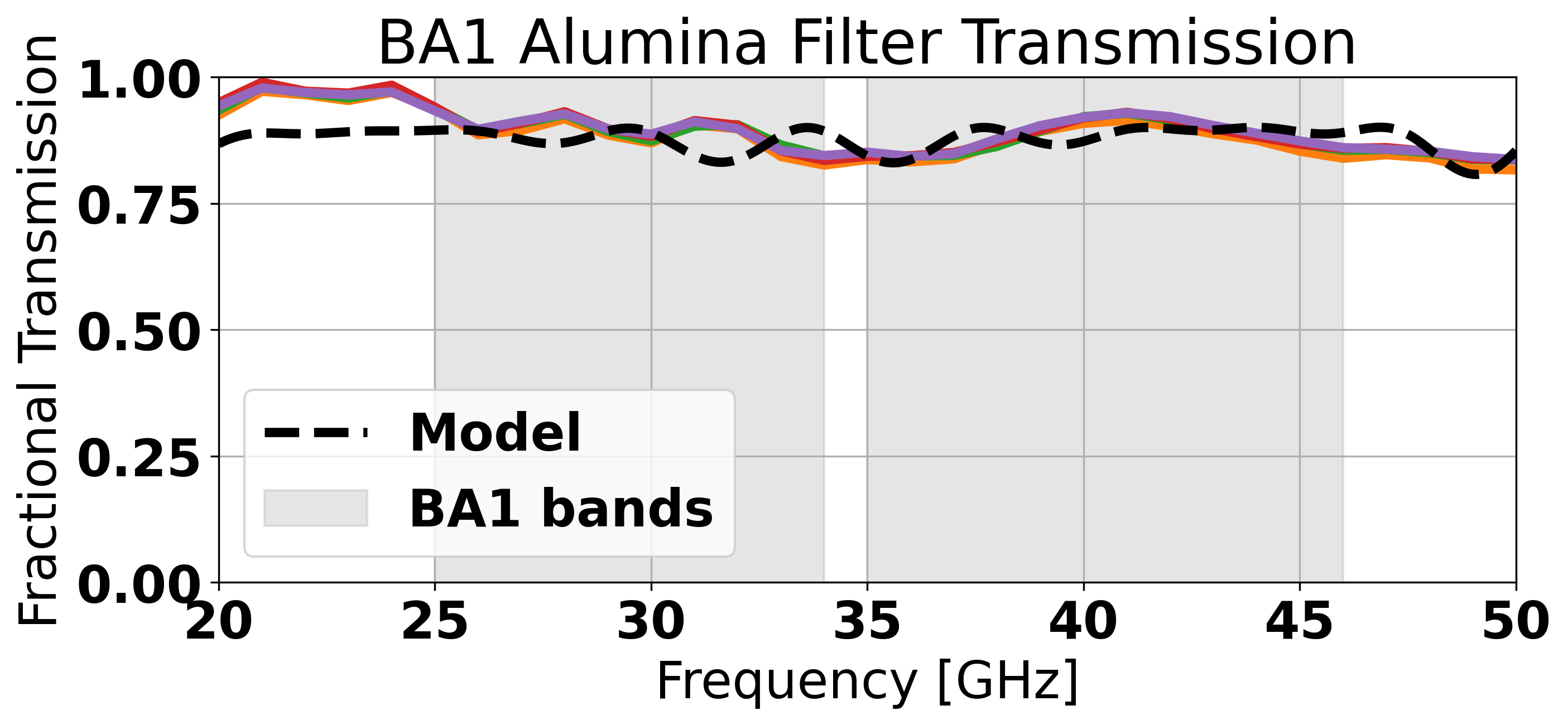}
\caption{Room-temperature transmission of the alumina infrared filter in the 30/40 GHz band of BA1. Colored lines show four consistent measurements taken with a Fourier Transform Spectrometer. The dashed line shows a model of the transmission, including room-temperature absorption (which decreases significantly at low temperatures). (Color figure online.)}
\label{fig1}
\end{center}
\end{figure}

With its high index of refraction, high thermal conductivity, low in-band loss and substantial IR blocking ability, aluminum oxide ceramic (also known as alumina) is a good candidate material for optics at millimeter wavelengths [4, 5]. A novel two-layer AR-coating process was developed for the first BICEP Array receiver alumina filter. Layers of Stycast 2850 and 1090 were precision-cast in a granite mold, using petrolatum and Sprayon{\textregistered} MR-314 as release layer. The sheets were then vacuum-bagged with the alumina filter for heat-lamination with low-density polyethylene (LDPE) as a bonding layer. To ensure durable adhesion after cryogenic cycling, the alumina was sandblasted with high-purity aluminum oxide, and after bonding to the alumina optics the AR layers were laser-diced for stress relief. The process resulted in a robust filter with excellent in-band transmission (see Fig.~\ref{fig1}). 

Antireflective layers for the subsequent BA receivers are currently in development following the same method. A single layer of mixed Stycast 2850 and 1090 is intended for the 95 and 150 GHz receiver alumina coatings. (The same mixture was used with a different, costlier application technique for BICEP3 [6].) Current development is focused on ensuring the thinner epoxy layers intended for higher frequencies reliably separate from the mold. The process is intended to work for curved lens surfaces as well, using mild heating ($40-50^\circ$~C) to deform the layers before bonding.

\section{Teflon AR-Coatings for Plastic Optics}
\label{sec_plastics}

HDPE and nylon optics are generally easier to manufacture than alumina and require lower-index antireflective layers. Two types of commercially available polytetrafluoroethylene (PTFE, also known as Teflon) membranes generally present densities and thicknesses suitable for plastic substractes at millimeter wavelengths: expanded PTFE (ePTFE) and sintered PTFE (sPTFE). Example formulations used in the BK program include Teadit$^{\textregistered}$\footnote{www.teadit.com} RGD ePTFE used in BA1, as well as Porex{\textregistered}\footnote{www.porex.com} sPTFE used in {\it Keck Array}. 

\begin{figure}[htbp]
	\begin{center}
		\includegraphics[width=0.8\linewidth, keepaspectratio]{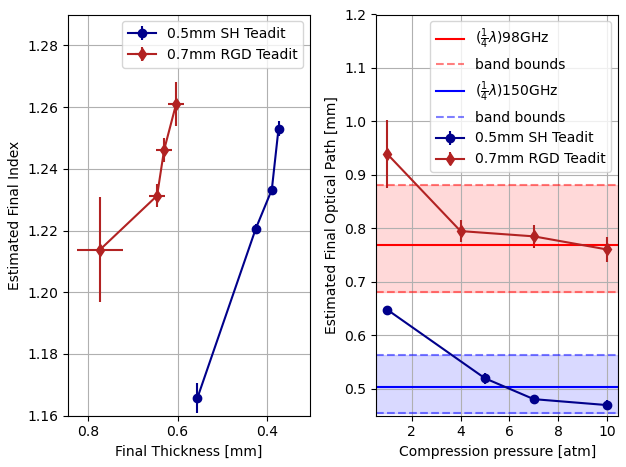}
		\caption{Optical properties of ePTFE samples heated to 137$^\circ$C and compressed at different pressures ranging from 1 to 10 atmospheres. {\it Left:} Colored lines show the increase in index and decrease in thickness of two ePTFE formulations. {\it Right:} Change in the optical path length for the same two ePTFE formulations treated at different pressures. Solid horizontal lines show targets for 95 GHz (red) and 150 GHz (blue). These results point to process pressures of $\sim$5-7 atmospheres for Teadit{\textregistered} RGD and SH in order to optimize performance in these two bands. (Color figure online.)}
			\label{fig2}
	\end{center}
\end{figure}

For future observations at 95~GHz and 150~GHz with BICEP Array-sized optics, none of the materials previously used for smaller-aperture receivers were available in sufficiently large sizes. As a result we developed a custom heat compression process to modify the thickness and refractive index of two ePTFE formulations: Teadit{\textregistered}~RGD and Teadit{\textregistered}~SH. As shown in Fig.~\ref{fig2}, we tuned the applied compression pressure in order to reduce the thickness and increase the density of $\sim$0.7~mm Teadit{\textregistered}~RGD and $\sim$0.5~mm Teadit{\textregistered}~SH gasket samples, to meet targets for quarter-wavelength AR-coatings at 95~GHz and 150~GHz, respectively. The compression set is performed at a temperature above the glass transition of PTFE [8], and is dimensionally stable under repeated vacuum and cryogenic cycling. For infrared filters made of nylon, which presents a higher index $n$ than HDPE, we found off-the-shelf sPTFE from Fluorseals{\textregistered} to be a good match for the desired AR index $\sqrt{n}$. Again the AR layers were heat-laminated to each optic using vacuum bagging and LDPE as a bonding layer. Differential thermal contraction between the plastic substrates and AR layers is minimal, resulting in excellent adhesion performance. 

\begin{figure}[htbp]
	\begin{center}
		\includegraphics[width=0.9\linewidth, keepaspectratio]{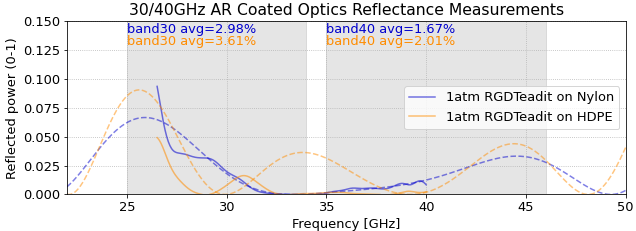}
		\includegraphics[width=0.9\linewidth, keepaspectratio]{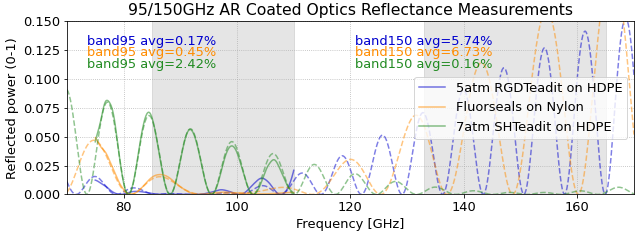}
		\caption{Room-temperature reflection spectra of AR-coated HDPE and nylon samples taken with a Vector Network Analyzer in two frequency bands (30/40 GHz and 95 GHz). Spectra measured at 95~GHz are extrapolated to 150~GHz to assess the performance of the Teadit{\textregistered} SH AR coating tuned for this band. Fit spectra are shown as dashed lines. The resulting average in-band reflections are very low for these AR- coatings designed for 30/40 GHz ({\it top}), 95 GHz ({\it bottom}, blue and orange lines) and 150 GHz ({\it bottom}, green line). (Color figure online.)}
			\label{fig3}
	\end{center}
\end{figure}

Fig.~\ref{fig3} presents reflectance measurements carried out with a Vector Network Analyzer at 30/40~GHz and 95~GHz. Off-the-shelf 1.3~mm Teadit{\textregistered}~RGD provided satisfactory performance for the BA1 receiver at 30/40~GHz. However, tuning the parameters of commercially available ePTFE membranes improved performance for 95~GHz and 150~GHz compared to the 30/40~GHz band. Table~\ref{table1} shows the final parameters of the AR layers for all three bands, as measured from fits to the reflectance spectra in Fig.~\ref{fig3}.

\begin{table}[]
	\centering
	\begin{tabular}{cc|cccc}
		 \textbf{Frequency} & \textbf{Substrate} & \textbf{AR material}   & \textbf{Compression} & \textbf{Index} & \textbf{Thickness (mm)}  \\ \hline
	 30/40 GHz & HDPE & Teadit{\textregistered} RGD ePTFE  & 1 atm & $1.231 \pm 0.002$ & $1.618 \pm 0.001$ \\
	 30/40 GHz & Nylon & Teadit{\textregistered} RGD ePTFE  & 1 atm & $1.226 \pm 0.001$ & $1.632 \pm 0.001$ \\
	 95 GHz & HDPE & Teadit{\textregistered} RGD ePTFE  & 5 atm & $1.243 \pm 0.005$ & $0.650 \pm 0.003$ \\
	 95 GHz & Nylon & Fluorseals{\textregistered} sPTFE  & 1 atm & $1.291 \pm 0.005$ & $0.580 \pm 0.002$ \\
	 150 GHz & HDPE & Teadit{\textregistered} SH ePTFE   & 7 atm & $1.25 \pm 0.01$ & $0.395 \pm 0.003$ \\
	\end{tabular}
	\caption{Final indices and thicknesses of AR layers designed for various plastic optics at 30/40~GHz, 95~GHz, and 150~GHz, fit from the reflectance spectra shown in Fig.~\ref{fig3}. These measurements show that the heat/compression process results in the desired changes in index and thickness of the Teadit{\textregistered} ePTFE.}
	\label{table1}
\end{table}

\section{Conclusions}

We have developed novel techniques to produce high-performance antireflective coatings for the alumina, HDPE and nylon optics in the 30/40~GHz, 95~GHz and 150~GHz bands. We will continue to validate their optical performance through reflectance and transmission measurements at 150~GHz. Finally, we will research new methods to produce the thin AR layers needed for the highest-frequency BICEP Array band at 220/270~GHz, as well as future receivers for the CMB-Stage 4 experiment.

\begin{acknowledgements}
	The BICEP/{\it Keck} projects have been made possible through a series of grants from the National Science Foundation including Grants No. 0742818, No. 0742592, No. 1044978, No. 1110087, No. 1145172, No. 1145143, No. 1145248, No. 1639040, No. 1638957, No. 1638978, and No. 1638970, and by the Keck Foundation. The development of antenna-coupled detector technology was supported by the JPL Research and Technology Development Fund, and by NASA Grants No. 06-ARPA206-0040, No. 10-SAT10-0017, No. 12- SAT12-0031, No. 14-SAT14-0009, and No. 16-SAT-16- 0002. The development and testing of focal planes were supported by the Gordon and Betty Moore Foundation at Caltech. Readout electronics were supported by a Canada Foundation for Innovation grant to UBC. Support for quasioptical filtering was provided by UK STFC Grant No. ST/N000706/1. The computations in this work were run on the Odyssey/Cannon cluster supported by the FAS Science Division Research Computing Group at Harvard University. The analysis effort at Stanford and S. L. A. C. is partially supported by the U.S. DOE Office of Science. We thank the staff of the U.S. Antarctic Program and in particular the South Pole Station without whose help this research would not have been possible. We thank all those who have contributed past efforts to the BICEP/{\it Keck} series of experiments, including the BICEP1 team. 
\end{acknowledgements}

\pagebreak

\end{document}